%% file: Manuscript_arXiv.tex
\documentclass[letterpaper]{article} 
\usepackage{aaai2026}  
\usepackage{times}  
\usepackage{helvet}  
\usepackage{courier}  
\usepackage[hyphens]{url}  
\usepackage{graphicx} 
\usepackage{natbib}  
\usepackage{caption} 
\usepackage{algorithm}
\usepackage{algorithmic}
\usepackage{newfloat}
\usepackage{listings}
\usepackage{outlines}
\usepackage{multirow}

\DeclareCaptionStyle{ruled}{labelfont=normalfont,labelsep=colon,strut=off} 
\floatstyle{ruled}
\newfloat{listing}{tb}{lst}{}
\floatname{listing}{Listing}
\AtBeginEnvironment{quote}{\itshape}

\begin{document}

\title{``Always Want to Use it for Everything'': Understanding Young Adults' Perceptions of AI Dependence}

\author{
    Ashlee Milton\textsuperscript{\rm 1},
    Leah Ajmani\textsuperscript{\rm 1}, 
    Amy Heger\textsuperscript{\rm 2}, 
    Forough Poursabzi-Sangdeh\textsuperscript{\rm 2}, 
    Mihaela Vorvoreanu\textsuperscript{\rm 2},
    Jina Suh\textsuperscript{\rm 3}
}
\affiliations{
    \textsuperscript{\rm 1}University of Minnesota\\
    \textsuperscript{\rm 2}Microsoft\\
    \textsuperscript{\rm 3}University of Washington\\
    \{milto064, ajman004\}@umn.edu, \{v-aheger, fpoursabzi, Mihaela.Vorvoreanu\}@microsoft.com, jinasuh@uw.edu
}

\maketitle

\begin{abstract}
The growing integration of general-purpose AI chatbots into people's daily lives has raised concerns about the potential for unhealthy dependence, particularly among young adults. As a first step toward understanding and characterizing AI chatbot dependence from the perspective of young adults, we collected testimonials from AI chatbot users aged 18 to 25 through an online questionnaire to capture their thoughts and experiences with this phenomenon. From participant responses, we identified three contributing factors of AI dependence: chronic use, efficiency, and delegation. The combination of these in a person's interaction behavior was considered to indicate AI dependence. Participants also observed feelings of atrophy in abilities from AI dependence, leading to psychological impacts such as feelings of inadequacy. Interpreting these findings through the lens of self-determination theory reveals how AI chatbot dependence can impact young adults' personal and social development. We argue that preventing lasting harm to young adults' development is paramount, and provide implications for rethinking AI chatbot dependence grounded in this understanding.
\end{abstract}

\input{Sections/Introduction}

\input{Sections/RW}

\input{Sections/Methods}

\input{Sections/Findings}

\input{Sections/Discussion}

\input{Sections/Conclusion}


\bibliography{Manuscript}
\clearpage
\appendix
\section{Additional Methods Details}
\label{app:methods}

\subsection{Additional Questionnaire \& Scale Information.}
We used and adopted several scales in our study questionnaire for investigating young adults perceptions of AI chatbot dependence. The scales we used are as follows:
\begin{itemize}
    \item Motivations: \textit{AI Use Motivation Scale}~\cite{huang2024ai,wolf2024chatgpt}
    \item Attitudes: \textit{General Attitudes towards Artificial Intelligence Scale}~\cite{schepman2020initial}
    \item Problematic Use: \textit{Problematic ChatGPT Use Scale}~\cite{yu2024development,zhou2024examining}
    \item Dependence: \textit{Affective Dependence Scale}~\cite{sirvent2022concept,zhou2024examining}, \textit{AI Dependence Scale}~\cite{huang2024ai}, and the \textit{Generative AI Addiction} measures from~\citet{zhou2024examining} 
\end{itemize}
The only changes we made to existing scales we used were to add the word "chatbot" after "AI" in the scale items for clarity, as well as removed overlapping items across scales and standardized Likert scales (respecting their original distributions) to reduce participant response fatigue. 

For investigating the influence of AI chatbot design on use, we created list of design features from prior work~\cite{flayelle2023taxonomy,zhou2024examining} on online platforms and dependence that have been shown to interact with internet or AI dependence or addiction. This resulted in 15 items that we asked how each item effected participants AI chatbot use from "significantly decreases" to "significantly increases" with a "not applicable" option for it the AI chatbot they used did not have a specific feature. 

\subsection{Response and Social Desirability Bias Mitigation Strategies.}
Going into this study, we were aware that the social conceptions and language surrounding AI chatbot dependence may influence how people respond to questions on the topic. With this in mind, we opted to use an anonymous online questionnaire to collect testimonials to help mitigate the social desirability effects. We also opted to use neutral and balanced language in our questions whenever possible. For example, when asking about about behaviors associated with AI chatbot dependence, we asked for positive and negative behaviors, noting that some dependence such as people being ''dependent'' on glasses to see, prompting participants to not only assume dependent behaviors were negative.

\section{Non-significant Results}
\label{app:results}
In this appendix section we provide Tables ~\ref{app:metrics} and ~\ref{app:assessments} that show all results that were not significant. As a reminder significance required that an item had a significant correlation with Likert-scale self-reported AI dependence and a significant difference between dependent and non-dependent participant groups (Mann-Whitney U of $p < 0.05$ with a Bonferroni correction and Cohen's D effect size of $>0.7$).

\begin{table}[h]
\centering
\resizebox{\linewidth}{!}{%
\begin{tabular}{|p{2.1cm}|p{5cm}|l|ll|}
\hline
                                             &                                                                               & Step 1:      & \multicolumn{2}{c|}{Step 2: Median}        \\ \cline{4-5} 
Metric                                       & Item                                                                         & Correlation   & \multicolumn{1}{l|}{D}    & ND   \\ \hline
\multirow{2}{2.1cm}{Time (7-point Scale)}    & Duration of Use                                                              & 0.21          & \multicolumn{1}{l|}{4} & 4 \\ \cline{2-5} 
                                             & Session Length of Use                                                        & 0.17          & \multicolumn{1}{l|}{2} & 2 \\ \hline
\multirow{15}{2.1cm}{Motivations (5-point Scale)} & I use AI chatbots to search for and obtain the information I need       & 0.30          & \multicolumn{1}{l|}{5} & 5 \\ \cline{2-5} 
                                             & I use AI chatbots to overcome certain difficulties                           & 0.34          & \multicolumn{1}{l|}{4} & 4 \\ \cline{2-5} 
                                             & I use AI chatbots to enhance my learning and knowledge                       & 0.26          & \multicolumn{1}{l|}{5} & 4 \\ \cline{2-5}
                                             & I use AI chatbots to entertain and relax myself                              & 0.22          & \multicolumn{1}{l|}{4} & 2 \\ \cline{2-5}
                                             & I use AI chatbots because it brings me happiness                             & 0.29          & \multicolumn{1}{l|}{3} & 2 \\ \cline{2-5}
                                             & I use AI chatbots because I need to interact with someone                    & 0.24          & \multicolumn{1}{l|}{2} & 1 \\ \cline{2-5}
                                             & I use AI chatbots because it helps me avoid feeling alone or lonely          & 0.27          & \multicolumn{1}{l|}{2} & 1 \\ \cline{2-5}
                                             & I use AI chatbots as a means of escaping from family, friends, stressful situations, or other problems & 0.15              & \multicolumn{1}{l|}{1} & 1 \\ \cline{2-5}
                                             & I use AI chatbots to forget about other things or avoid responsibilities     & 0.22          & \multicolumn{1}{l|}{2} & 1 \\ \cline{2-5}
                                             & I use AI chatbots because I am encouraged or required to                     & 0.24          & \multicolumn{1}{l|}{2} & 2 \\ \cline{2-5}
                                             & I use AI chatbots because I am told not to use it                            & 0.11          & \multicolumn{1}{l|}{1} & 1 \\ \cline{2-5}
                                             & I use AI chatbots because many people I know use it                          & 0.15          & \multicolumn{1}{l|}{2} & 2 \\ \cline{2-5}
                                             & I use AI chatbots because people I respect use it                            & 0.29          & \multicolumn{1}{l|}{2} & 1 \\ \cline{2-5}
                                             & I use AI chatbots to look cool                                               & 0.21          & \multicolumn{1}{l|}{1} & 1 \\ \cline{2-5}
                                             & I use AI chatbots for help accomplishing tasks more quickly                  & 0.26          & \multicolumn{1}{l|}{5} & 5 \\ \cline{2-5}
                                             & I use AI chatbots to improve my performance or productivity                  & 0.34          & \multicolumn{1}{l|}{5} & 4 \\ \hline
\multirow{14}{2.1cm}{System Design (5-point Scale)} & How quickly the AI chatbot responds to me                             & 0.18          & \multicolumn{1}{l|}{5} & 4 \\ \cline{2-5}
                                             & How predictable the AI chatbot's responses are                               & 0.15          & \multicolumn{1}{l|}{3} & 3 \\ \cline{2-5}
                                             & How often the AI chatbot reminds me to interact with it                      & 0.33          & \multicolumn{1}{l|}{3} & 2 \\ \cline{2-5}
                                             & How the AI chatbot influences my decisions and actions                       & 0.25          & \multicolumn{1}{l|}{4} & 3 \\ \cline{2-5}
                                             & How consistently available and accessible the AI chatbot is                  & 0.19          & \multicolumn{1}{l|}{5} & 4 \\ \cline{2-5}
                                             & How I can verbally speak to the AI chatbot, and it speaks back               & 0.19          & \multicolumn{1}{l|}{4} & 3 \\ \cline{2-5}
                                             & How the AI chatbot adapts to my wants and needs                              & 0.27          & \multicolumn{1}{l|}{5} & 4 \\ \cline{2-5}
                                             & How the AI chatbot remembers me or past conversations                        & 0.24          & \multicolumn{1}{l|}{5} & 4 \\ \cline{2-5}
                                             & How I can personalize the AI chatbot to my specifications                    & 0.22          & \multicolumn{1}{l|}{4} & 4 \\ \cline{2-5}
                                             & How the AI chatbot looks like a person                                       & 0.20          & \multicolumn{1}{l|}{3} & 3 \\ \cline{2-5}
                                             & How the AI chatbot behaves like a person                                     & 0.31          & \multicolumn{1}{l|}{4} & 3 \\ \cline{2-5}
                                             & How the AI chatbot portrays sentience or consciousness                       & 0.23          & \multicolumn{1}{l|}{4} & 3 \\ \cline{2-5}
                                             & How much the AI chatbot costs                                                & 0.16          & \multicolumn{1}{l|}{3} & 3 \\ \cline{2-5}
                                             & How knowledgeable or intelligent the AI chatbot is                           & 0.28          & \multicolumn{1}{l|}{5} & 5 \\ \hline
\end{tabular}%
}
\caption{Non-significant AI chatbot usage item correlations and medians for dependent (D) and non-dependent (ND) participants.}
\label{app:metrics}
\end{table}

\begin{table}[h]
\centering
\resizebox{\linewidth}{!}{%
\begin{tabular}{|p{3cm}|p{5cm}|l|ll|}
\hline
                                                               &                                                                                         & Step 1:     & \multicolumn{2}{c|}{Step 2: Median}        \\ \cline{4-5} 
Scale                                                          & Item                                                                                    & Correlations & \multicolumn{1}{l|}{D}    & ND   \\ \hline
\multirow{14}{3cm}{Attitudes Towards AI Chatbots (5-point Scale)} & For regular transactions, I would rather interact with an AI chatbot than with a human & 0.28 & \multicolumn{1}{l|}{2} & 2 \\ \cline{2-5} 
                                                               & AI chatbots can provide new economic opportunities for my country                       & 0.27  & \multicolumn{1}{l|}{4} & 3 \\ \cline{2-5}
                                                               & AI chatbots can help people feel happier                                                & 0.26  & \multicolumn{1}{l|}{4} & 4 \\ \cline{2-5} 
                                                               & I am impressed by what AI chatbots can do                                               & 0.35  & \multicolumn{1}{l|}{5} & 4 \\ \cline{2-5}
                                                               & AI chatbots can have a positive impact on people's wellbeing                            & 0.28  & \multicolumn{1}{l|}{4} & 4 \\ \cline{2-5}
                                                               & AI chatbots can perform better than humans                                              & 0.33  & \multicolumn{1}{l|}{4} & 2.5 \\ \cline{2-5}
                                                               & Organizations use AI chatbots unethically                                               & -0.24 & \multicolumn{1}{l|}{3} & 4 \\ \cline{2-5} 
                                                               & I think AI chatbots make many errors                                                    & -0.23 & \multicolumn{1}{l|}{4} & 4 \\ \cline{2-5}
                                                               & I find AI chatbots sinister                                                             & -0.25 & \multicolumn{1}{l|}{2} & 2 \\ \cline{2-5}
                                                               & AI chatbots might take control of people                                                & -0.07 & \multicolumn{1}{l|}{2} & 2 \\ \cline{2-5}
                                                               & I think AI chatbots are dangerous                                                       & -0.26 & \multicolumn{1}{l|}{2} & 3 \\ \cline{2-5}
                                                               & I shiver with discomfort when I think about future uses of AI chatbots                  & -0.22 & \multicolumn{1}{l|}{2} & 3 \\ \cline{2-5}
                                                               & People like me will suffer if AI chatbots were used more and more                       & -0.24 & \multicolumn{1}{l|}{2} & 3 \\ \cline{2-5}
                                                               & AI chatbots are used to spy on people                                                   & -0.03 & \multicolumn{1}{l|}{3} & 3 \\ \hline
\multirow{7}{3cm}{Problematic Use (4-point Scale)}             & I constantly have thoughts related to AI chatbots lingering in my mind                  &  0.30 & \multicolumn{1}{l|}{1} & 1 \\ \cline{2-5}
                                                               & I frequently find myself opening AI chatbots even when I had no initial intention to use one & 0.37 & \multicolumn{1}{l|}{2} & 1 \\ \cline{2-5}
                                                               & I have lost interest in previously enjoyable activities due to using AI chatbots        & 0.19 & \multicolumn{1}{l|}{1} & 1 \\ \cline{2-5}
                                                               & My use of AI chatbots causes me to procrastinate and delay completing necessary tasks   & 0.28 & \multicolumn{1}{l|}{2} & 1 \\ \cline{2-5}
                                                               & I suffer from sleep deprivation due to excessive use of AI chatbots                     & 0.27 & \multicolumn{1}{l|}{1} & 1 \\ \cline{2-5}
                                                               & I hide the extent of my AI chatbot usage from family, friends, or therapists            & 0.23 & \multicolumn{1}{l|}{1} & 1 \\ \cline{2-5}
                                                               & I turn to AI chatbots to alleviate feelings of helplessness or anxiety                  & 0.34 & \multicolumn{1}{l|}{2} & 1 \\ \hline
\multirow{2}{3cm}{Emotional Dependence (5-point Scale)}        & When an AI chatbot distances itself from me I feel an unbearable emptiness              & 0.34 & \multicolumn{1}{l|}{1} & 1 \\ \cline{2-5} 
                                                               & I would like to make friends with AI chatbots                                           & 0.34 & \multicolumn{1}{l|}{2} & 1 \\ \hline
\multirow{4}{3cm}{AI Dependence \& Addiction (4-point Scale)}  & The use of AI chatbots has damaged my relationships                                     & 0.20 & \multicolumn{1}{l|}{1} & 1 \\ \cline{2-5} 
                                                               & I have tried to reduce the amount of time I spend on AI chatbots but failed             & 0.27 & \multicolumn{1}{l|}{1} & 1 \\ \cline{2-5}
                                                               & Using AI chatbots sometimes interferes with my other activities (e.g. school, work, social) & 0.30 & \multicolumn{1}{l|}{1} & 1 \\ \cline{2-5}
                                                               & I find it difficult to control my use of AI chatbots                                    & 0.34 & \multicolumn{1}{l|}{1} & 1 \\ \hline
\end{tabular}%
}
\caption{Non-significant AI dependence measures item correlations and medians for dependent (D) and non-dependent (ND) participants.}
\label{app:assessments}
\end{table}

\end{document}

%% file: Sections/Introduction.tex
\section{Introduction}
General-purpose AI chatbots powered by large language models are being rapidly integrated into people's everyday lives. The number of adult AI chatbot users has doubled since 2023~\cite{sidoti202534}. With increased adoption of AI chatbots, researchers and popular media are raising concerns about the potential cognitive effects this technology may have on users. These concerns include effects on mental health~\cite{fang2025ai,iftikhar2025llm}, social relationships~\cite{Sanford2026mix}, and cognitive abilities~\cite{grinschgl2022supporting,Hogenboom2026stupid}. Concerns about the toll AI chatbot use takes on users are particularly salient for young adult users, who are not only among the fastest adopters~\cite{sidoti202534}, but also in vulnerable transitional states~\cite{patrick2020patterns}. 

One fundamental framing for understanding the potential impacts on users is reliance-based, focusing on concepts such as overreliance, dependence, and addiction. In extreme cases, prior work found addiction-like behaviors in some young users~\cite{namvarpour2026understanding,xie2023friend}. For AI chatbot dependence, a growing body of literature translates symptoms and behaviors of substance use disorder to AI chatbot use~\cite{kooli2025generative,shen2025dark,huang2025exploring,salah2026me}. However, there is a challenge in that addiction and dependence are situated in a clinical context. Therefore, these translation efforts risk over-pathologizing certain AI chatbot use behaviors as addiction~\cite{ciudad2025people}.

Generally, AI dependence is defined as the excessive use of AI technologies that leads to dependence and addictive trends~\cite{huang2024ai}, based largely on clinical perspectives~\cite{naseer2025psychological,zhai2024effects}. While researchers continue to uncover evidence of AI chatbots' impact on users, there remains a lack of understanding of how people who use AI chatbots perceive and experience AI dependence. Assessments grounded in an accurate portrayal of AI dependence are important because these measures rely on users' self-reports~\cite{schepman2020initial,yu2024development,sirvent2022concept}. Specifically, the lacking understanding of how young people express their experience of AI dependence could contribute to missing valuable indicators that do not align with existing definitions. Our study asks: \emph{how do young adult users of AI chatbots perceive and experience AI chatbot dependence, and how do these understandings align with current conceptions and measures of AI dependence?}

To better understand AI chatbot dependence (AI dependence, for short) from a lived-experience perspective, we gathered testimonials from general-purpose AI chatbot users aged 18--25. We used an online questionnaire with open-ended items, along with questions on AI chatbot use and validated dependence measures \cite[e.g.,][]{schepman2020initial,yu2024development,sirvent2022concept}. Our findings show that participants who self-identified as dependent on AI chatbots did not score as dependent on existing AI dependence measures, suggesting a misalignment between these measures and self-reported AI dependence. To understand this discrepancy and illuminate our participants' perceptions of AI dependence, we conducted a qualitative analysis of their testimonials. We use quantitative results to contextualize and corroborate the qualitative findings. Participants' understanding of AI dependence was based on three contributing factors: \textit{chronic use}, \textit{efficiency}, and \textit{delegation}. They describe how these factors could indicate dependence on AI chatbots and discuss downstream impacts, primarily from atrophy of ability and loss of confidence.

To consider the implications of these findings, we use the lens of Self-Determination Theory (SDT)~\cite{deci2012self,ryan2000self}, which posits autonomy, competence, and relatedness as basic psychological needs for human well-being. Through this lens, the factors participants identify as contributing to AI dependence can negatively impact these needs. Chronic use and delegation to AI chatbots can reduce young adults' autonomy. Atrophy of ability can diminish competence. Aspects of AI chatbot use for socioemotional purposes point to issues of relatedness. When these basic psychological needs are unmet, the personal growth of a generation of young adults is at risk. As such, our work highlights the need to rethink how AI dependence is conceptualized and provides implications for policy, design, and societal norms surrounding AI dependence and its risks. 

%% file: Sections/RW.tex
\section{Related Work}
\subsection{Dependence and Addiction to Technology}
The risks of technology use, specifically the concept of technology-based dependence, have been documented in prior research across smartphones, the internet, and social media. Nearly two decades ago, mobile phone dependence was linked to impulsivity and problematic behavior among users~\cite{billieux2007does}. Then, with increased internet access, scholars saw the rise of internet addiction, which is associated with psychological symptoms, such as obsessive-compulsiveness, depression, anxiety, paranoid ideation, and psychoticism~\cite{adalier2012relationship}.

Most relevant to our focus on AI chatbots is the body of literature on the psychological impacts of social media. Scholars have found heavy social media use to be associated with social comparison and depressive symptoms~\cite{feinstein2013negative}, social media fatigue~\cite{dhir2018online}, and social isolation~\cite{whaite2018social}. More recent analyses have shown that the motivation to use social media to maintain social relationships is associated with increased loneliness~\cite{bonsaksen2023associations}, highlighting that technology use for socioemotional goals can have paradoxical effects. 

While many of these findings may point to technology dependence and addiction being a user problem, scholars have documented ``dark design patterns''~\cite{mildner2023engaging,shi2026siren} or specific platform design patterns that are causally implicated. Past research has demonstrated that vulnerability to technology-specific dependence is not uniform across individuals. For example, ~\citet{chak2004shyness} has shown that shyness, loneliness, and external locus of control are predictors of internet addiction. Similarly, \citet{epley2008creating} explained the mechanism linking loneliness to dependence as anthropomorphism, or treating nonhuman agents as humanlike. There is mounting evidence that the technologies themselves play a major role in dependence and addiction issues. ~\citet{flayelle2023taxonomy} conducted a review of various online technology design features that influence users' control over engagement and use, thereby promoting addictive behaviors. These features include variable ratio reinforcement schedules, social rewards and validation, and continuous information renewal~\cite{flayelle2023taxonomy,orben2024mechanisms}. Prior work on technologies designed to engage users cognitively and socially set the stage for studying the risks posed by current AI chatbots and AI chatbot dependence~\cite{mahari2025addictive}.

\subsection{Psychosocial and Cognitive Impacts of AI Chatbots}
Scholars have reported that AI chatbot use can lead to a sense of short-term relief for users, but simultaneously raise concerns about long-term safety and effectiveness~\cite{song2025typing,siddals2024happened}. AI chatbots can enable avoidance or shift perceived social costs (i.e., burdensomeness) associated with human connections~\cite{ajmani2025seeking}. AI chatbot use duration has been found to predict worse loneliness, more emotional dependence, and more problematic use. High usage and social attraction towards AI chatbots were correlated with self-reported dependence~\cite{fang2025ai}. Long-term use, on the other hand, was found to significantly increase users' perceived attachment to AI chatbots and perceived empathy, while interpersonal relationships were largely maintained~\cite{chandra2025longitudinal}. However, there are still concerns about substituting AI chatbots for human connection, as such substitution may lead to AI dependence. One study suggests that the key predictors of the development of problematic use patterns may be loneliness and social anxiety~\cite{hu2023social}. Dependence signals, such as increased use of AI chatbots as a primary mental health resource and the development of emotional attachment to AI chatbots, point to the need to study dependence even in non-substitutive contexts.

The effects of AI chatbot use are not limited to socioemotional aspects. There are also implications for skill atrophy and self-efficacy. AI chatbot use can impair cognitive abilities~\cite{zhai2024effects}, particularly surrounding understanding and learning~\cite{shen2026ai}, as well as reduced critical thinking~\cite{gerlich2025ai,lee2025impact}. Users also experience negative changes to self-perception and agency when AI chatbots outperform them~\cite{schaap2024outperformed}. The influence AI chatbots have over users' cognitive abilities also has implications on their perceived autonomy~\cite{catena2026ai}. 

Given the effects of AI chatbots on users, researchers have begun to explore how specific AI chatbot design features propagate these harms. Harmful design patterns in AI companions have been found to include the absence of natural relationship endpoints, high attachment anxiety, and propensity to engender protectiveness~\cite{knox2025harmful}, with a taxonomy of the resulting harms being developed~\cite{zhang2025dark}. In some cases, AI chatbot design fails to account for the various ways users may engage with them, such as for mental health support, and thus fails to set appropriate boundaries~\cite{iftikhar2025llm}.

\subsection{AI Chatbot Dependence and Addiction Framing}
A relevant starting point for studying potential psychological effects of AI chatbots are framings of addiction and dependence. Several works have applied frameworks of dependence, often grounded in addiction literature, to users' interactions with AI chatbots. For example, ~\citet{zhou2024examining} engages with the idea of AI chatbot addiction by examining AI chatbot interactions using the Cognitive-Affective-Conative (CAC) theoretical framework. Regarding AI chatbot dependence, Goodman's behavioral dependence framework and self-determination theory have been used to develop a Generative AI Dependency Scale~\cite{goh2025generative}. In the last few years, several scales to measure AI chatbot dependence have been developed. ~\citet{yu2024development} created the Problematic ChatGPT Use Scale specifically for ChatGPT use and captures absorption and problematic engagement behaviors. Other scales have considered AI dependence in relation to users' motivational patterns~\cite{huang2024ai} and the emotional and relational aspects of AI chatbot use~\cite{sirvent2022concept}. While prior work demonstrates a proliferation of AI dependence and addiction framing, these works focus mostly on the clinical and measurement aspects of AI dependence, rather than the everyday experiential aspects.

Recent work has started to consider people's experiences with AI chatbot dependence, noting perceived psychological risks. From discussions of AI chatbot use on Reddit, five experiential dimensions of AI chatbot risks were identified, including difficulties with self-regulation and personalization associated with emotional dependence~\cite{zhu2026understanding}. Other literature \cite[e.g.,][]{chandra2025lived,zhang2025dark} uses lived experience methodology to conceptualize psychological risks that clinical inventories might miss.

Prior work shows that AI dependence is real: having been hypothesized from early technology-specific dependence research and observed in recent empirical studies. However, measurement and conceptualization of AI dependence mainly originate from clinical frameworks. Thus, in this work, we aim to build an AI dependence understanding from users' perceptions and experiences. We focus on young adults due to their vulnerability~\cite{patrick2020patterns} and fast adoption of AI chatbots~\cite{sidoti202534}.

%% file: Sections/Methods.tex
\section{Methods}
We collected testimonials from 290 young adults (aged 18 to 25) who regularly interact with AI chatbots via an online questionnaire. Participants were asked about their use of AI chatbots and how they conceptualize AI chatbot dependence. The study was approved by the ethics review board at Microsoft. In this section, we cover information about participants, the questionnaire, and the analysis methodology.

\subsection{Study Sample \& Recruitment}
Our study specifically focuses on how young adults conceptualize AI dependence. We chose to focus on this population because young adults are at transitional points in their lives~\cite{patrick2020patterns}, where they may turn to technology for support, and are among the fastest adopters of AI chatbots~\cite{sidoti202534}. To target young adults, we chose two main recruitment venues: an internal intern newsletter and Prolific, which allows for age-based recruitment. We collected data between June and August 2025.

Prolific participants completed a pre-screening questionnaire that verified individuals' eligibility, recorded their consent to participate, and collected demographic information before being invited to participate in the study. The study questionnaire verified and collected the same information from all participants, regardless of the recruitment method. 

Our inclusion criteria specified that participants must be 18 to 25 years old, currently reside in the US, and use AI chatbots at least once a week. Prolific participants had an additional criterion: a 100\% approval rate and at least 20 previous task submissions. In total, 290 participants consented to the study and completed the questionnaire. Participants were compensated with a \$20 digital gift card, or in the case of Prolific, with \$0.40 for the screening questionnaire and \$9 for the full questionnaire. The difference in compensation was due to Prolific's compensation norms and guidelines, which differ from the norms for other recruitment avenues.

\begin{table}[h]
\centering
\resizebox{0.9\linewidth}{!}{%
\begin{tabular}{|r|l|r|l|}
\hline
\multicolumn{1}{|l|}{\textbf{Demographic Variables}} & \textbf{N}               & \multicolumn{1}{l|}{\textbf{Demographic Variables}} & \textbf{N}               \\ \hline
\textbf{Age}                     &     & \textbf{Gender}                      &     \\ \hline
18-19                            & 27  & Man                                  & 127 \\
20-21                            & 77  & Non-Binary/Non-Conforming            & 14  \\
22-23                            & 81  & Transgender                          & 14  \\
24-25                            & 105 & Women                                & 153 \\ \cline{1-2}
\textbf{Ethnicity}           &  & Prefer not to answer                                & 1                        \\  \hline
Asian                                                & 81                       & \textbf{Education}          & \\ \cline{3-4}
Black                            & 43  & Associate Degree/Vocational Training & 26  \\ 
Hispanic                         & 46  & Bachelor's Degree                    & 103 \\
Indigenous                       & 11  & Doctorate Degree                     & 1   \\
Middle Easter/North African      & 3   & High School Graduate                 & 53  \\
Native Hawaiian/Pacific Islander & 2   & Master's Degree                      & 21  \\
White                            & 151 & Some College, No Degree              & 85  \\
Prefer not to answer             & 5   & Prefer not to answer                 & 1 \\ \hline
\end{tabular}%
}
\caption{Aggregated demographic information of participants.}
\label{tab:demo}
\end{table}

Our participants were on average 22 years old (M=22.3), with the majority identifying as women (n=157), white (n=154), and having a bachelor's degree (n=104). Table~\ref{tab:demo} overviews participants' demographics. Not all demographic counts align with the number of participants as we allowed multiple selection of many demographic questions. 

\subsection{Questionnaire}
Our online questionnaire had four parts: (1) Demographics, (2) AI usage, (3) AI dependence, and (4) Assessments. Demographics included questions about age, gender, ethnicity, and education. For AI usage, we asked closed-ended questions about time-based use, motivations for using AI~\cite{huang2024ai,wolf2024chatgpt}, and the influence of AI chatbots' design on use~\cite{flayelle2023taxonomy,zhou2024examining} to add context to participants' testimonials. The AI dependence section gathered participants' testimonials about their perspectives and experiences with AI dependence through open-ended questions about definitions of AI dependence and associated behaviors. We also asked closed-ended questions about whether they identified with those definitions or knew someone who did. For those who identified as feeling dependent on AI chatbots or knowing someone who was, we asked further questions about specific behaviors that indicated AI dependence. Lastly, we asked closed-response questions based on items from current measures of AI dependence~\cite{schepman2020initial,yu2024development,zhou2024examining,huang2024ai,sirvent2022concept}. For more details on the scales used in the questionnaire, please refer to Appendix~\ref{app:methods}.

\subsection{Analysis}
To analyze the open-response questions in the questionnaire, we used qualitative inductive thematic analysis~\cite{clarke2015thematic}. Two research team members each independently coded a set of related questions to distill initial themes. These themes were then discussed with the whole team in an iterative process until final themes was established and agreed upon. These resulting themes centered around the contributing factors and perceived consequences of AI chatbot dependence, which are reported in Section~\ref{sec:findings}. The final themes were consistent across all participant responses, regardless of whether they experienced AI dependence, observed someone with AI dependence, or neither.

While we analyze all participant responses, we use quotes exclusively from participants who either identified as feeling dependent or knew someone they believed was dependent on AI chatbots in our findings. We do this to center these users' lived experiences and to reflect the overarching themes that emerged around behaviors and experiences (both personal and observed) related to AI dependence. Participants are identified as either a ``D'' for self-reported dependent participants' responses or a ``B'' for bystander participants' responses regarding others' AI dependence.

While the main focus of this work is participants' open-ended responses to their lived experiences, we also analyze closed-ended responses to provide context for their experiences. We use a two-step process to conduct our analysis:
\begin{itemize}
    \item Step 1: We first the correlations between closed-ended items and participants' self-reported dependence, measured on a 1 ("strongly disagree") to 5 ("strongly agree") Likert scale. Only items with significant Spearman's rank correlation coefficients ($p < 0.05$) were included in step 2. We use Spearman's correlation as it is commonly used for ordinal data, such as Likert scales~\cite{schober2018correlation}. Given the number of correlations, we used a Bonferroni correction ($\alpha$ = 0.05; $m$ = 80, the number of individual correlations to self-reported dependence computed) to account for possible Type I errors. 
    \item Step 2: We separated participants into two groups. Participants who responded with either ``somewhat agree'' or ``strongly agree'' to the self-report AI dependence question were labeled dependent (n=74), and the remaining participants were considered non-dependent (n=216). We then tested for differences between the dependent and non-dependent participant groups using the Mann-Whitney U test, given the ordinal nature of the data, and report group medians. Differences between groups are only considered significant if $p < 0.05$ and Cohen's D effect size is $>0.7$. There were also a number of group comparisons, so we once again used a Bonferroni correction ($\alpha$ = 0.05; $m$ = 70, the number of group comparisons) to account for possible Type I errors.
\end{itemize}

%% file: Sections/Findings.tex
\section{Findings}
\label{sec:findings}
In investigating participants' understanding of AI chatbot dependence, we found a misalignment between current conceptions of AI dependence and our participants' perceptions and experiences. Participants' self-reported dependence did not align with current notions of AI dependence in existing measures. To understand this misalignment, we focused on participants' testimonies to reveal how they understand AI dependence and how it aligns with their reported use. 

We first present quantitative results based on participants' responses to closed-ended questions about their AI use and responses to AI dependence measures. We then share qualitative findings from participants' testimonies about their perceptions and experiences of AI dependence, using the quantitative results to contextualize and corroborate participants' understanding where appropriate.

\subsection{Quantitative Results: AI Chatbot Use and Dependence Measures}
In this section, we present the statistical analysis of participants' responses to AI chatbot use and AI dependence measures. We report results only when items have a significant correlation with Likert-scale self-reported AI dependence (step 1) and a significant difference between dependent and non-dependent participant groups (step 2). These findings focus primarily on group differences. Overall, we find significant differences in participants' frequency of AI chatbot use, motivations for reducing effort, and the use of validation by AI chatbots. Furthermore, we note a misalignment between current AI dependence measures and our participants' self-reported AI dependence.

For understanding participants' responses, we use the plain-language associated with Likert scale scores. We do not provide a breakdown of group demographic information, as we found no significant differences. Correlation coefficients with self-reported dependence and group medians for significant results are shown in Tables~\ref{tab:metrics} (usage) and~\ref {tab:assessments} (measures). Non-significant results are reported in Appendix~\ref{app:results}. 

\begin{table}
\centering
\resizebox{\linewidth}{!}{%
\begin{tabular}{|p{2.1cm}|p{5cm}|l|ll|}
\hline
                                             &                                                                               & Step 1:             & \multicolumn{2}{c|}{Step 2: Median}        \\ \cline{4-5} 
Metric                                       & Item                                                                          & Correlation & \multicolumn{1}{l|}{D}    & ND   \\ \hline
\multirow{2}{2.1cm}{Time (7-point Scale)}        & Frequency of Use                                                          & 0.53 & \multicolumn{1}{l|}{5} & 3 \\ \cline{2-5} 
                                             & Change Over Time (COT) of Use                                                 & 0.35 & \multicolumn{1}{l|}{5} & 4 \\ \hline
\multirow{4}{2.1cm}{Motivations (5-point Scale)} & I use AI chatbots to help me and make my life easier                      & 0.30 & \multicolumn{1}{l|}{5} & 4 \\ \cline{2-5} 
                                             & I use AI chatbots to improve my abilities                                     & 0.37 & \multicolumn{1}{l|}{5} & 4 \\ \cline{2-5} 
                                             & I use AI chatbots to manage my feelings                                       & 0.35 & \multicolumn{1}{l|}{2} & 1 \\ \cline{2-5} 
                                             & I use AI chatbots for the pleasure it gives me to know more about the chatbot & 0.34 & \multicolumn{1}{l|}{3} & 1 \\ \hline
System Design (5-point Scale)                & How the AI chatbot validates or supports my thoughts and feelings             & 0.35 & \multicolumn{1}{l|}{4} & 3 \\ \hline
\end{tabular}%
}
\caption{AI chatbot usage item correlations and medians for dependent (D) and non-dependent (ND) participants.}
\label{tab:metrics}
\end{table}

\subsubsection{Time-based Use}
We asked participants to report on the time spent using AI chatbots from four different perspectives: time since first use (Duration), change in their use over time (COT), frequency of use (Frequency), and length of use in a single setting (Session).

Contrary to current legislation on AI chatbot companions, which focuses on longer sessions as an indicator of problematic use, we find that only Frequency and COT show significant differences between groups. Dependent participants reported using AI chatbots ``several times a day'', with non-dependent participants reporting use at ``several times a week''. We see less of a difference between groups for change in use over time, with dependent participants' use having ``increased a lot'', and non-dependent participants' use ``increased a little''.

Our participants reported behaviors that suggest that how often users turn to AI chatbots and changes in their behavior over time may be more useful indicators of possible AI dependence than session length. 

\subsubsection{Use Motivations}
We considered participants' motivations for using AI chatbots. Motivations fell into two categories: external and internal. Each category had 10 motivations, with one external and three internal motivations showing significant relationships with AI dependence.

The only significant external motivation was, ``I use AI chatbots to help me and make my life easier''. While both groups reported agreement with this statement, dependent participants ``strongly agreed'' and non-dependent participants ``somewhat agreed''. Related to this idea of ``easier lives'', we saw internal motivation to ``use AI chatbots to improve my abilities'' linked to feelings of dependence. Dependent participants ``strongly agreed'' with this motivation, while non-dependent participants only ``somewhat agreed''. 

We also saw two internal motivations related to emotions correlating with AI dependence. Regarding motivations ``to manage my feelings'', non-dependent participants ``strongly disagreed'' with the concept, whereas dependent participants ``somewhat disagreed''. A similar pattern emerged across groups for the motivation of ``the pleasure it gives me to know more about the chatbot'', with dependent participants being more ``neutral'' to non-dependent participants strong disagreement. It is worth noting that only five dependent participants reported using AI chatbots for companionship-related reasons, such as talking to favorite characters. The interpretation of this last motivation may be the perspective of pleasurable interactions rather than relationship building.

Considering the motivations our dependent participants agreed they had for using AI chatbots, making life easier and improving abilities, we see a theme related to reducing effort. If something is easier or an ability is augmented, the effort required to accomplish it is reduced. We can even draw a connection to managing feelings, as it shares the emotional burden, thus reducing the effort needed to carry it.

\subsubsection{Influence of System Design on Use}
To examine how AI chatbot design may influence AI dependence, participants were asked how various design aspects of AI chatbots affected their use. Across 15 possible design aspects, only one showed a significant impact between groups: validation. Participants were asked to consider the statement ``how the AI chatbot validates or supports my thoughts and feelings'' regarding its effect on their use of AI chatbots. We saw that for dependent participants, validation ``somewhat increased'' their use, while non-dependent participants noted it had a closer to ``no effect'' on their use.

Thinking about validation alongside participants' increased motivation to manage emotions and seek pleasurable interactions paints a potentially concerning picture of how AI chatbot design may be interacting with participants' emotional states and contributing to AI dependence. Remember that very few of our participants used AI chatbots for companionship. Thus, we are seeing emotional effects in more general-purpose uses of AI chatbots.

\subsubsection{Relationship Between AI Chatbot Dependence Measures and Self-Reports}
We examine the relationship between existing measures and self-reported AI dependence to understand how these measures correlate with personal perceptions of AI dependence. See Table~\ref{tab:assessments} for all significant statistics. AI dependence measures are designed to be administered as self-reported behaviors and cover people's attitudes, use, and dependence on AI chatbots. In examining these relationships, we begin to see a connection between participants' attitudes, reliance, and emotional consequences. However, our participants, on average, disagreed with items from current scales intended to measure AI dependence.

\begin{table}
\centering
\resizebox{\linewidth}{!}{%
\begin{tabular}{|p{3cm}|p{5cm}|l|ll|}
\hline
                                                               &                                                                                         & Step 1:     & \multicolumn{2}{c|}{Step 2: Median}        \\ \cline{4-5} 
Scale                                                          & Item                                                                                    & Correlations & \multicolumn{1}{l|}{D}    & ND   \\ \hline
\multirow{6}{3cm}{Attitudes Towards AI Chatbots (5-point Scale)} & There are many beneficial applications of AI chatbots                                 & 0.35        & \multicolumn{1}{l|}{5} & 4 \\ \cline{2-5} 
                                                               & I am interested in using AI chatbots in my daily life                                   & 0.50        & \multicolumn{1}{l|}{4} & 3 \\ \cline{2-5} 
                                                               & AI chatbots are exciting                                                                & 0.44        & \multicolumn{1}{l|}{4.5} & 4 \\ \cline{2-5} 
                                                               & Much of society will benefit from a future full of AI chatbots                          & 0.44        & \multicolumn{1}{l|}{4} & 3 \\ \cline{2-5} 
                                                               & I would like to use AI chatbots in my own job                                           & 0.45        & \multicolumn{1}{l|}{4} & 3 \\ \cline{2-5} 
                                                               & An AI chatbot agent would be better than an employee in many regular jobs               & 0.32        & \multicolumn{1}{l|}{3} & 2 \\ \hline
Problematic Use (4-point Scale)                                & I become fully absorbed when using AI chatbot                                           & 0.46        & \multicolumn{1}{l|}{3} & 1 \\ \hline
\multirow{4}{3cm}{Emotional Dependence (5-point Scale)}        & Honestly, I always need AI chatbots to be available                                     & 0.50        & \multicolumn{1}{l|}{2.5} & 1 \\ \cline{2-5} 
                                                               & I honestly believe that if I lost access to AI chatbots, I would not be able to bear it & 0.45        & \multicolumn{1}{l|}{1} & 1 \\ \cline{2-5} 
                                                               & I sincerely believe that I need AI chatbots more than others                            & 0.45        & \multicolumn{1}{l|}{2} & 1 \\ \cline{2-5} 
                                                               & I think I am emotionally dependent on AI chatbots                                       & 0.39        & \multicolumn{1}{l|}{1} & 1 \\ \hline
\multirow{2}{3cm}{AI Dependence \& Addiction (4-point Scale)}  & I rely too much on AI chatbots                                                          & 0.44        & \multicolumn{1}{l|}{2} & 1 \\ \cline{2-5} 
                                                               & I feel uneasy, anxious, or upset when I cannot use AI chatbots                          & 0.46        & \multicolumn{1}{l|}{2} & 1 \\ \hline
\end{tabular}%
}
\caption{AI dependence measures item correlations and medians for dependent (D) and non-dependent (ND) participants.}
\label{tab:assessments}
\end{table}

When considering participants' attitudes towards AI chatbots from the \textit{General Attitudes towards Artificial Intelligence Scale}~\cite{schepman2020initial}, we find that those who identified as dependent had significantly more positive views overall than those who were not. Looking at differences across individual items, themes emerge around the benefits and job use of AI chatbots. Dependent participants had more positive responses to items about finding ``AI chatbots exciting'' and about the ``beneficial applications'' or ``benefits from a future full of AI chatbots'' compared to non-dependent participants. Furthermore, they felt more positively towards using ``AI chatbots in their daily lives'' and ``in their own jobs'' than non-dependent participants.

For the remaining assessments, we report findings by item themes rather than by overall assessment, as overall scores on all the AI dependence assessments were on the ``disagree'' side of the scale for both groups.

The most prominent theme among items with significant differences across scales was reliance. We can see this directly from a significant item on the \textit{AI Dependence Scale}~\cite{huang2024ai,zhou2024examining}: ``I rely too much on AI chatbots''. In more nuanced ways, we see this concept in items related to needing AI chatbots ``more than others'' or ``always be available'', from the \textit{Affective Dependence Scale}~\cite{sirvent2022concept,zhou2024examining}, and ``I become fully absorbed when using AI chatbots'' from the \textit{Problematic ChatGPT Use Scale}~\cite{yu2024development,zhou2024examining}. 

We see possible emotional consequences of AI dependence from scale items. Items like feeling unable to ``bear the loss'' of access (\textit{D-Mdn}=1, \textit{ND-Mdn}=1, $p<.0001$), or feeling ``emotionally dependent'' on AI chatbots (\textit{D-Mdn}=1, \textit{ND-Mdn}=1, $p<.0001$), from the \textit{Affective Dependence Scale} are examples, although they show no difference in median. Furthermore, the item ``I feel uneasy, anxious, or upset when I cannot use AI chatbots'' from the \textit{AI Dependence Scale} speaks directly to possible emotional consequences.

Even with these emerging themes, remember that our dependent participants, on average, disagreed with the items on the dependence assessment scale. This suggests that current assessments do not align with our participants' behaviors or conceptualizations of AI dependence. 

Our analysis shows that participants' self-reported behaviors do not align with some current conceptions of AI dependence, particularly regarding time-based use and dependence measures. The qualitative results shed light on the sources of this misalignment, providing an understanding of AI dependence grounded in lived experience.

\subsection{Qualitative Results: User-Centered Understanding of AI Dependence}
In this section, we present participants' understanding of AI dependence. Participants noted three factors they perceived as contributing to AI dependence: \textit{chronic use}, \textit{efficiency}, and \textit{delegation}. Additionally, they highlighted feelings of ability atrophy that led to psychological effects, such as loss of confidence, which they associated with AI dependence. Together, these contributing factors and atrophy of ability were central to how they understood and perceived AI dependence. Themes in this section were consistent across all participants, including those who experience AI dependence (dependent participants, noted by D\#) and others' observations of those they consider dependent on AI chatbots (bystander participants, noted as B\#). Where relevant, we include references from our analysis of AI use and dependence measures to contextualize themes. 

\subsubsection{Contributing Factors of AI Chatbot Dependence}
\paragraph{Chronic Use.} The first factor of AI dependence concerns participants' chronic use of AI chatbots, specifically along dimensions of (1) habitual use and (2) acontextual use. Habitual use was characterized by frequent use of AI chatbots as a necessary resource, while acontextual use was observed across many contexts, regardless of appropriateness.

\textit{1) Habitual Use.} When discussing their use of AI chatbots, dependent participants used the word ``frequently'' or its synonyms in their responses. It is common for participants' statements to start with ``I use [AI chatbots] frequently..'' (D183) or ``I constantly talk to the chatbot..'' (D153). Bystander participants also noted the frequency with which dependent users accessed AI chatbots, who observed that these users ``bring it up frequently in real-life conversation`` (B234) or how ``every time [they] have a conversation, [dependent user] has to pause to look up answers and information using AI'' (B207).

Participants further described needing to turn to AI chatbots as a first resource in many situations. Dependent participants echoed D239, who said ``AI is the first thing [they] think of when [they] have a question''. In some cases, participants were ``not able to initiate projects without brainstorming with AI'' to the point of ``habitually'' using it (D133). Participants noted feeling ``slightly distressed'' (D45) when AI chatbots were unavailable to ``double check with ... to reassure [themselves]'' (D72). Bystander participants noticed how dependent users ``always have ChatGPT open'' or ``keep hearing how this task can be done by a chatbot'' (B35).

\textit{2) Acontextual Use.} Acontextual use describes how dependent users view AI chatbots as universally useful tools, regardless of context. D162 stated that they ``like to use it for a lot of things, such as making a plan, or fact-checking information''. The view that AI chatbots could be used in many contexts was prevalent among dependent participants. However, regular use of AI for ``quick information, writing support, decision-making help, and planning tasks'' also made them feel dependent (D134).

Bystanders noticed dependent users treating AI chatbots as a ``go-to solution for everything'', noting being told to just ``ask ChatGPT'' when interacting with dependent users (B201). The lack of resource diversity was also not lost on bystander participants, who recognized how dependent users ``will use AI for every question they have instead of googling and scouring through sites'' (B253). Overall, bystander participants shared a similar sentiment to B130, who said that dependent users ``always want to use it for everything''.

\textit{Connecting to Quantitative Results:} Themes of habitual and acontexual use echo findings from participants' AI chatbot use in the quantitative results. Recall that dependent participants reported significantly more frequent use of AI chatbots (see ``Time-based Use''), which aligned with participants' perceptions of habitual use. The connection to reliance and the emotional consequences of AI dependence is apparent in participants' feelings of a pervasive need to use AI chatbots, leading to psychological distress when they are unavailable (see ``Relationship Between AI Chatbot Dependence Measures and Self-Reports'').

\paragraph{Efficiency.} Efficiency for our participants involved using AI chatbots to increase productivity or free up time. From participants' responses, efficiency moved beyond traditional work productivity into personal aspects of life, but was not directly associated with offloading complex processes in cognitive or socioemotional contexts.

Participants focused on using AI chatbots for efficiency of ``more mundane or repetitive'' (D1) tasks, to complete them ``quickly and effectively'' (D97). D40 shared that ``finding information with AI chatbots has been more efficient'', implying efficiency reflected positively ``on [their] work and learning performance''. However, participants noted a trade-off: they felt more efficient, but their \textit{role}, or position relative to the process, had changed. For example, D115 said they were not ``doing [their] own research, deciding to save time and use AI chatbots to make [their] life easier''. D30 directly recognized a shift in their role as a programmer, shifting from creator to reviewer, from the drive for efficiency:
\begin{quote}
    ``[I am] using it to write large parts of code when, before chatbots, I would be writing it all myself. I have also noticed a pressure in the world to move faster, and spending hours unaided on a problem has a low return on investment.''
\end{quote}

We also see a drive for efficiency in personal aspects of people's lives, such as health and emotional support. Participants found that they ``don't want to take the time to research an exercise routine'' or ``skincare routine'' and instead ``have ChatGPT do it'' (D229). This efficiency extended to tasks many would usually turn to other people for, like emotional support. D187 mentioned that they turned to AI chatbots to ``help me breakdown relationship thoughts at any point in the day''. Others, like D79, ``use AI for support when no humans are available'' or to avoid bothering others.

\textit{Connecting to Quantitative Results:} Participants' perceptions of efficiency corroborate our quantitative results on self-reported attitudes and motivations. Recall that a primary motivator of AI dependence was to make life easier (see ``Use Motivations''). This concept appears here, with the added context of how it affects participants' roles in their lives and interactions with others. Findings around the role AI chatbots have in daily life also aligns with the more positive attitudes seen in dependent participants responses on attitudes towards AI chatbots (see ``Relationship Between AI Chatbot Dependence Measures and Self-Reports''). Taken together, these findings of putting less effort into one's life and acceptance of AI chatbots paint a picture of AI dependence effecting the role people have in their own lives.

\paragraph{Delegation.} In examining participants' responses, we notice a trend: dependent users delegate some of their cognitive processes, such as (1) learning, (2) decision-making, and (3) socioemotional processes, to AI chatbots. We see that participants identified such delegations as a factor contributing to AI dependence, noting gaps developing in the cognitive processes of dependent users.

\textit{1) Learning.} Participants' accounts of school and interpersonal communication suggest delegation of their learning to AI chatbots. The goal of many dependent participants' delegation of schoolwork was to pass their classes rather than to learn. D196 described how, for their ``math class, [they] completely depended on AI to help [them] pass''. Participants noted they are ``getting better grades'' (D106, D148) when delegating their learning to AI chatbots, reinforcing their behavior. Delegation of interpersonal skill development was also common as dependent participants felt AI chatbots' ``allowed [them] to be more social and communicate better by giving [them] constructive feedback'' (D164). 

In delegating learning, the roles of dependent participants shifted once again as they no longer actively engaged in learning but instead focused on simply getting an answer, with as little resistance as possible. We can see this in D114's recounting of their feelings of dependence:
\begin{quote}
    ``I feel ... somewhat dependent on [AI chatbots], as ... I know that this information will not help me further down the road. So I use the AI chatbot to just get it over with.''
\end{quote}
The impact of this outcome-driven approach from AI chatbot dependence is directly stated by D106: ``having AI do [their] schoolwork for [them], instead of doing it [themselves], has decreased the amount of learning I've done’’. From an outside perspective, participants like B285 felt that dependent users were ``making no effort to learn, grow, or think independently''. 

\textit{2) Decision-Making.} Delegation extends beyond learning to general decision-making. Using AI chatbots for life decisions was common in participants' responses. Examples from dependent participants ranged from basic decisions, such as ``figuring out where to eat at a local restaurant'' (D79), to important ones, such as ``work, health, and life decisions'' (D129). Participants felt that delegating decision-making could ``help [them] in making the right decisions'' (D247). Dependent participants like D116 even struggled to make decisions on their own as they felt ``like [they] need to talk to a chatbot before [they] make a decision.''

The decision-making role is shifted from people to AI chatbots, which bystander participants expressed concerns over. Bystander participants noticed how people ``immediately seek “counsel” with the [AI chabot] or about their decisions, big or small'' (B285). Using AI chatbots for big decisions was seen as a sign of struggle:
\begin{quote}
    ``I know that this person has asked [AI chatbots] to come up with a daily routine for their children, including when they should get up and go to bed. The fact that they are even using it to raise their children is indicative to me that they are struggling... '' (B118)
\end{quote}
The shifting role in decision-making for dependent users was seen to cause their personal development to stall or take a ``catastrophic nosedive'' (B231, B69). 

\textit{3) Socioemotional Processes.} Participants also discussed delegating social and emotional cognitive processes. AI chatbots were used to ``help console friends and family'' (D13) or for ``guidance on how to handle [their] relationship'' (D225). Bystander participants reported hating it when AI chatbots were used to talk to them, stating ``if [they] wanted to talk to a robot, [they] would'' instead wanting ``an actual human response'' (B201). Bystander participants noted they felt that ``people's relationships take a toll because they are consulting people less'' (B138) and ``neglecting their relationships'' (B245).

On an emotional level, dependent participants delegated their self-soothing to AI chatbots. Participants would ``vent out [their] frustrations about personal struggles'' but noticed ``an unhealthy habit of using it for mental and emotional health'' (D221). Using AI chabots to ``discover [their] emotional state'' (D187)  or ``help [them] sort through emotions'' (D61), to the point where ``losing it would be some detriment to [them]'' (D183). In some cases, dependent participants ended up turning to AI chatbots ``over consulting people [they] know in real life'' (D31) or feeling ``like [they] have to talk to a chatbot to feel better ... if [they are] having a big emotion such as anxiety'' (D116). Bystander participants speculated that dependent users ``had better times talking to an AI chatbot`` due to receiving ``validation from the chatbot'', thus not feel the need to talk to others (B193).

While the factors of Efficiency and Delegation are related, participants framed them differently. Participants used efficiency to refer to usage of AI mainly to save time and reduce effort on general tasks, which, in turn, shifted their role in the process (e.g., from creator to reviewer). Delegation, on the other hand, involved handing over the cognitive processes themselves (learning, decision-making, and socio-emotional work) to AI, going beyond a mere role change and leaving gaps in the their own abilities.

\textit{Connecting to Quantitative Results:} Shifting roles resulting from cognitive process delegation can be contextualized in our quantitative results on dependent participants' motivations for using AI chatbots. Recall that dependent participants' motivations for using AI chatbots revealed that they were more motivated than others to improve their abilities and manage their emotions (see ``Use Motivations''). These results corroborate current themes in the delegation of cognitive processes, ultimately leading dependent users to exert less effort on these processes and allowing AI chatbots to take on the primary role. Additionally, we saw that validation in system design is related to AI dependence (see ``Influence of System Design on Use''), supporting participants' suspicions of its role in AI dependence.

\subsubsection{Perceived Consequences of AI Dependence: Feelings of Ability Atrophy}
Above, we describe the factors that participants perceived as contributors to AI dependence. Below, we articulate the perceived effects of this dependence: feelings of abilities atrophying, leading to psychological impacts such as loss of confidence. We find these feelings of atrophy in both experiences among dependent participants and in bystander observations.

\paragraph{Feelings of Atrophy.} Participants described feeling their abilities, or cognitive processes, atrophying through subtle changes in their behavior or mindsets. D30 references atrophy in how they ``have noticed [their] coding skills begin to atrophy as I spend less time coding and more time reviewing code''. Not all dependent participants use the word atrophy, but still noted feelings of ``skills  [being] less exercised'' (D9) due to them resorting ``to AI chatbots too quickly, even though [they] know that it is probably good for [them] to do the work [themselves]'' (D24). Bystander participants also perceived dependent users not ``developing real-world skills'' and ``turning to AI chatbots as a crutch'' (B284).

A common cognitive process that participants felt was atrophying is critical thinking. There were many mentions by bystander participants of how dependent users were ``car[ing] less about their own original thinking'' (B24) and not ``stopping and trying to think themselves'' (B44). Dependent participants mentioned how they ``immediately ask AI for help before actually thinking about the solution`` (D3). This atrophy is clear in responses like D155's, who said:
\begin{quote}
    ``I use it to completely do my assignments for me instead of using it to help me learn things. I barely even read through the questions; I just used the answers it gave me instead of having to think for myself.''
\end{quote}
Participants recognized that the atrophy of critical thinking could have long-term implications, such as career stagnation, if they did not ``stop relying on [AI chatbots]" and ``start thinking for [themselves]'' (D72).

As dependent users turn to AI chatbots for communication and emotional support instead of other people, bystander participants recognized that social abilities were also subject to atrophy. Participants noted that some dependent users felt they were ``not able to communicate with someone without asking a chatbot what to speak'' (B25) or were unable to ``provide a response or feedback until they ask AI`` (B107). In the case of feelings of atrophy for managing emotions, one bystander participant put it aptly when they said: 
\begin{quote}
    ``I think [their] mental well-being has been affected negatively because [they] cannot process [their] emotions without the help of AI.''
\end{quote}

\paragraph{Psychological Effects of Atrophy.} The feelings of dependent participants' abilities atrophying had a real effect on them psychologically. From dependent participants' responses, we see them expressing feelings of inadequacy and impostor syndrome, which negatively affect their mental well-being. Experiences of questioning their own abilities and feelings of inadequacy were common among dependent participants. D260 noted that they began to think ``maybe I don't actually have as many skills as I think I do and just have false security from my chatbot usage''. These thoughts lead to questions about their ability to succeed without an AI chatbot, as seen in D31's response:
\begin{quote}
    ``I almost feel helpless sometimes without an AI chatbot: even though logically I know that I am capable of learning new things without a chatbot, it feels much more intimidating to tackle a very new problem without it.''
\end{quote}

We see a tension in dependent participants' responses between knowing they are capable people and feeling lost without AI chatbots, which affects their self-confidence and leads them to feel like impostors. Bystander participants noted how dependent users had an `inability to trust [themselves] made [them] second-guess [themselves]'' and relied on AI chatbots ``for an extra sense of security'' (B51). Feelings of self-doubt appear to bystanders as dependent users feeling ``like an impostor'' and ``nervous to ask questions in real-life'' (B6), ultimately losing ``the confidence to solve problems or manage emotions independently'' (B284).

%% file: Sections/Discussion.tex
\section{Discussion}
In this work, we set out to build an understanding of AI chatbot dependence grounded in young adults' lived experiences. Drawing on participants' self-reported behaviors, perceptions and beliefs captured through an online questionnaire, our findings illuminate the factors participants identify as contributing to AI chatbot dependence and how they relate to self-reported AI usage. We further highlight the felt and observed connection between the atrophy of users' abilities and AI chatbot dependence. 

We apply self-determination theory (SDT) to discuss the implications of our work. SDT posits that people have three basic psychological needs: autonomy, competence, and relatedness. All three are essential to a person's growth and well-being~\cite{deci2012self,ryan2000self}. These basic needs are also core components in positive computing~\cite{calvo2014positive,peters2020tools}, which aims to bring well-being psychology to technology design. SDT has been previously applied in studies of AI chatbots~\cite{nguyen2018understanding}, mainly in the domain of learning~\cite{xia2023mediating,annamalai2023exploring,li2025systematics} or AI chatbot design~\cite{sims2026ai,yang2021designing}. Thus, SDT is a relevant interpretive lens in discussing the harms of dependence on general-purpose AI chatbots among young adults. 

Given our goal of understanding AI dependence from young adults' perspective, we provide implications for rethinking AI dependence. We focus on implications for improving AI dependence measures, policy, and design, as well as consideration of weighing social pressures and AI use. 
 
\subsection{Self-Determination Theory and AI Dependence}
SDT states that an individual's personal growth and mental well-being are harmed when their basic needs are unmet. For young adults at pivotal points in their lives, this stagnation of personal growth can have long-term implications. Here, we draw direct connections between the diminishing of the three basic psychological needs from SDT and our findings. 

\subsubsection{Autonomy}
Our findings reveal a lack of ownership over actions from our participants. Across all three contributing factors (efficiency, delegation, and chronic use), our findings highlight a common issue around autonomy (i.e., a person's feeling of control and endorsement of their behaviors~\cite{deci2012self}). We find that human autonomy decreases when AI chatbots are \textbf{chronically used} (which, in our findings, comprises of habitual and acontextual use) to \textbf{delegate} aspects of users' lives in the name of \textbf{efficiency}. This dynamic aligns with prior work about how AI chatbots may erode autonomy~\cite{krook2025autonomy}, highlighting behavior such as anthropomorphism as a potential contributing factor~\cite{marchegiani2025anthropomorphism,peter2025benefits}.

We also confirm recent evidence on AI chatbots' influence on users' autonomy, particularly in decision-making~\cite{catena2026ai}. Participants reported shifting their decision-making to AI chatbots, removing a sense of ownership over important life decisions. In other words, AI chatbot dependence fundamentally shifts a user's role in their own lives, making them passive observers rather than active owners. As a user's autonomy or ownership of their actions is reduced, a lack of accountability for those actions often follows~\cite{witvliet2023accountability}, making it difficult for people to take responsibility. When users lose autonomy and accountability, they stop growing as people, as indicated by our dependent participants. Our findings, therefore, raise awareness of a negative reinforcing effect of the loss of autonomy on users' growth and the subsequent development of dependence.

\subsubsection{Competence}
We find that dependence is highly associated with feelings of ability atrophy, which mirrors the concept of competence in SDT. Competence is one's experience of effectiveness in action~\cite{deci2012self}. Participants talked about not exercising their skills and feeling them atrophy as a result. Our findings connect these actions to the perception of ability atrophy and a lack of confidence in their ability to work independently, which, in turn, can promote their continued dependence on AI chatbots.

These findings echo concerns that AI chatbots may affect users' cognitive functions, specifically regarding how cognitive offloading can cause atrophy~\cite{dergaa2024tools}. Recent work has examined the effects of AI chatbots on people's cognitive abilities, revealing that AI tool usage affects critical thinking~\cite{gerlich2025ai,lee2025impact} and overreliance on AI chatbots negatively impacts cognitive abilities~\cite{zhai2024effects} and causes negative self-perception~\cite{schaap2024outperformed}. In other words, dependence on AI chatbots may stifle the development of an ability by diminishing competence, discouraging practice, and harming growth.

\subsubsection{Relatedness}
Across our findings, we saw dependent users tending to either not interact with people or use AI chatbots to interact with others, affecting those relationships. These findings point to SDT's need for relatedness, i.e., the need to feel connected with others~\cite{deci2012self}. AI dependence may undermine relatedness by fundamentally changing when and how users interact with others. 

Creating and maintaining relationships requires effort~\cite{yong2025m,canary2006maintaining}. However, as AI chatbots aim to provide frictionless interactions through constant availability and validation, they create unrealistic expectations for social interactions. Our participants expressed that leveraging AI chatbots instead of genuinely interacting with other people reduces the effort they put into those relationships, causing them not to establish new ones or to strain existing ones. These findings align with recent work on sycophantic AI, showing that AI chatbots can give users a false sense of correctness or moral endorsement~\cite{cheng_elephant_2025}; lead to reduced willingness to take action and repair interpersonal conflicts~\cite{cheng_sycophantic_2025}; and interact with human cognitive biases to produce feedback loops that progressively reinforce maladaptive beliefs in vulnerable users and increase social isolation~\cite{dohnany2026feedbackloops}. Thus, through frictionless design, AI chatbots may inadvertently facilitate the distancing of AI-dependent young adults from others, affecting their social and emotional development. 

By viewing young adults' understanding of AI dependence through the lens of SDT, we show how their conceptualization of AI dependence relates to autonomy, competence, and relatedness, and highlight its potential effects on personal growth and well-being. Next, we discuss what our findings mean for reconceptualizing AI dependence.

\subsection{Rethinking AI Dependence}
The goal of this work was to understand how young adults perceive and experience AI chatbot dependence. In doing so, we identified implications for reconceptualizing AI dependence and its measurement, as well as for the policies and designs of AI chatbots. We end our discussion with considerations for balancing current societal pressures surrounding AI chatbot use with awareness of AI dependence risks.

\subsubsection{Implications for AI Reconceptualization}
Our quantitative results show that our self-reported dependent participants did not consistently score as ``dependent'' on AI dependence measures from previous work~\cite{schepman2020initial,huang2024ai,zhou2024examining}. This misalignment between our participants' perceptions of AI dependence and current assessments suggests that solely measuring AI dependence in the same way as other forms, such as substance dependence, may not be appropriate. Therefore, we call for adjusting the conceptualization and measurement of AI dependence. Measuring AI dependence must start with AI users' lived experiences and their own feelings of dependence, especially when assessments rely on self-reporting. Our findings suggest several key factors in the experience of AI dependence that should be considered in future measurements, including chronic use, efficiency, delegation, feelings of ability atrophy, loss of confidence, and effects on social connection. Future research could explore quantifying and testing these concepts to develop updated psychometric approaches to AI dependence.

\subsubsection{Implications for Policy and Design}
Our study's findings also point to specific design and policy considerations for addressing usage patterns related to AI dependent.

First, our findings suggest expanding our current understanding of time spent using AI chatbots. Current legislation in some US states requires warnings based on the latter; users receive a warning after three hours of use for ``companion chatbots''~\cite{CASB243,NYArticle47}. In our findings, AI dependence is correlated with how frequently users turn to AI, not how long they use it in a single session. While the present findings should not be interpreted as sufficient grounds for immediate policy reform, they suggest that frequency of use may deserve consideration in future policy discussions.

Second, we join other researchers and practitioners in calling out the importance of reducing ``sycophantic'' behavior and indiscriminate validation in AI. Similar to previous work, which highlights the negative impact of such AI behavior on users' prosocial tendencies~\cite{cheng2026sycophantic}, we find that participants would rather speak to a chatbot than a human, partly because of AI's validating behaviors. Unlike human-human interactions, human-AI interactions create a frictionless communication environment where users hold unilateral power. As a result, users may lose their capacity to manage the socioemotional discomfort inherent to human communication. Further research is needed to identify the best design approaches for reducing validation language in AI outputs to preserve the skills necessary to meet users' needs for relatedness.

\subsubsection{Considerations for Balancing Societal Pressures}
While AI chatbots can be helpful, the risks of AI dependence and its consequences underscore the need for a measured approach. Societal pressures emphasizing the importance of AI use and the utmost value of efficiency should be counter-balanced with information and support about the risks of AI chatbot use, such as dependence. By highlighting key characteristics of AI dependence and sharing participants' experiences, we hope this work can alert young adults and society at large to the risks of AI dependence, enabling them to identify it early and to take steps to mitigate its impact. However, it is important to keep in mind the role society plays in AI chatbot adoption, influencing it through social norms, media influence, and the fear of missing out~\cite{yang2025technological,li2024cannot}. AI education and literacy efforts alone might not be sufficient to counterbalance the current societal zeitgeist that strongly favors the use of AI chatbots. As a society, we need to consider what a balanced, sustainable approach to AI use might look like and work towards it at a pace that matches the speed of AI development.

\subsection{Limitations}
We note several limitations of the current work. While we make general claims about the population of young adults, our sample is not representative of all young adults or their experiences. Our study was conducted in the United States and did not consider potential cultural factors that may influence perceptions of AI. Future work should consider a more diverse cultural context to understand young adults' experiences with AI dependence. Methodologically, our participants' testimonies were collected through short-answer responses in an online questionnaire, which, while allowing anonymity, may not provide the same level of nuance as in-depth interviews. Similarly, we relied on self-reported AI use behaviors, which may be subject to response and social desirability biases (see Appendix~\ref{app:methods} for more information on out mitigation strategy). Taken together, these limitations suggest that our findings offer a valuable but necessarily partial view of AI chatbot dependence. While lived experience is fundamental to understanding the phenomenon of AI dependence in young adults, it is important to complement the perspective our participants shared with others', inclusive of other populations and methodological approaches.

%% file: Sections/Conclusion.tex
\section{Conclusion}
In this work, we built an understand of AI chatbot dependence from young adults' perspective. Through an online questionnaire of general-purpose AI chatbot users, we identified three contributing factors of AI dependence (chronic use, efficiency, and delegation) from participants' perceptions and experiences. Furthermore, we found dependent users perceive an atrophy of their abilities, leading to feelings of inadequacy. We leverage self-determination theory to highlight the implications of AI dependence on young adults' personal and social development. We contribute to the understanding AI dependence from young adults perspective, highlighting negative effects on their development, and implications for rethinking AI dependence to prevent lasting harms. Future research building on this work can pursue updating AI dependence assessments that better capture the phenomenon as currently experienced by young adults.